# Biased Nanoscale Contact as Active Element for Electrically Driven Plasmonic Nanoantenna

Alexander V. Uskov*[1,2], Jacob B. Khurgin[3], Mickael Buret[4], Alexandre Bouhelier[4], Igor V. Smetanin[1], and Igor E. Protsenko[1]

[1] P. N. Lebedev Physical Institute, Leninsky pr. 53, 119991 Moscow, Russia

[2] ITMO University, Kronverkskiy pr. 49, 197101 Sankt-Petersburg, Russia

[3] Department of Electrical & Computer Engineering, John Hopkins University, Baltimore, Maryland, USA

[4] Laboratoire Interdisciplinaire Carnot de Bourgogne, CNRS-UMR 6303, Université Bourgogne Franche-Comté, 21078 Dijon, France

*Supporting Information*

**ABSTRACT:** *Electrically-driven optical antennas* can serve as compact sources of electromagnetic radiation operating at optical frequencies. In the most widely explored configurations, the radiation is generated by electrons *tunneling* between metallic parts of the structure when a bias voltage is applied across the tunneling gap. Rather than relying on an inherently inefficient inelastic light emission in the gap, we suggest to use a ballistic nanoconstriction as the *feed element* of an optical antenna supporting plasmonic modes. We discuss the underlying mechanisms responsible for the optical emission, and show that with such a nanoscale contact, one can reach quantum efficiency orders of magnitude larger than with standard light-emitting tunneling structures.

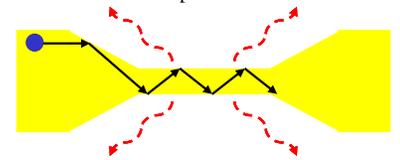

KEYWORDS: *excitation of plasmons, inelastic tunneling, mesoscopic contacts, nanoconstriction, ballistic transport, ballistic transport*

The development of *electrically-driven plasmonic nanoantennas* is gaining a general interest in the field of nanoplasmonics[1]. The attractive feature of electron-fed nanoantennas is the unprecedented co-integration of the electrical drive and the source of radiation in a single nanoscale geometry that can be integrated in emerging plasmonic nanocircuitries (plasmonic nanochips)[2]. The nanoantennas are essentially constituted of metals and do not incorporate semiconductor materials, which facilitates their miniaturization. Another benefit of using metal units is the fast to ultrafast dynamics and expected bandwidth. Unlike semiconducting optical sources limited by intrinsic and extrinsic parameters (carrier diffusion, nonradiative decay channels, etc), the modulation bandwidth of electron-fed optical antennas may reach and extend beyond the THz regime, thus offering a possibility to create unique ultrasmall and ultrafast optoelectronic devices.

By default, the term nanoantenna denotes an electrically-contacted metal nanostructure with a nanometer-size *gap*. When a bias voltage is applied across the gap, electrons tunnel between the metallic parts of the structure, and may couple to electromagnetic oscillations mediated by the radiative decay of surface plasmon polaritons (SPPs). Excitation of SPPs with electron *inelastic* tunneling is far from being a new topic – early studies can be traced back to the pioneering 1976 work by Lambe and McCarthy on light emission from planar Metal-Insulator-Metal (MIM) structures[3] and was further discussed in the context of light emission observed at the tip of a Scanning Tunneling Microscope (STM).[4] The quest for an all-electrical source of SPPs had revived the interest in excitation of SPPs with tunneling electrons is recent years[5-15]. Thanks to that body of literature, there is a consensus on the underlying physics of SPP excitation in MIM structures (see [16-17] and references therein). With this understanding, the SPP is excited in the process of inelastic tunneling hence it is reasonable to refer to this process as "SPP-assisted tunneling". Unfortunately, this process competes with a very strong elastic tunneling across the gap, hence MIM structures suffer from low quantum efficiency (QE ~ $10^{-4} - 10^{-5}$) of photon emission, even in the presence of a strong Purcell enhancement. It is however informative to compare this external efficiency with other types of nanoscale sources. Upon absorption of a photon, quantum dots and molecules are typically engineered to provide a high probability of fluorescence emission; some internal quantum efficiencies are approaching unity. Considering a diffraction-limited focal spot, the external efficiency is on the order of $10^5$ incoming photons to produce a fluorescence photon. For electrically-pumped devices such as electroluminescent carbon nanotube (CNT) transistors, the typical operating currents are in the μA regime, indicating that approximately $10^{12}$ carriers are injected in the device every second. The efficiency of CNT radiative recombination was estimated to be around $10^{-6}$ photons/recombination under ambipolar operation[18] and a two to three orders of magnitude increase in efficiency for defect-induced luminescence was observed[19], a conversion rate comparable to photoluminescence efficiencies discussed above for common nanoscale source of visible light.

Fig. 1a illustrates the mechanism of SPPs excitation in a tunneling structure. When the bias voltage $V$ is applied between two metals separated by the insulator, electrons may tunnel from one metal to other *elastically* (green arrow), i.e., without loss of energy, or *inelastically* (red arrow), with emission of the quantum $\hbar\omega_{sp}$ into a plasmonic mode of the structure. Note that the elastically tunneled electrons transfer their extra energy to cold electrons in electron-electron (e-e) collisions, so that the electron temperature may increase up to several thousand K causing the black body-like radiation observed in [7,15] in addition to the SPP excitation in the gap. The main reason for low efficiency of SPP emission in a MIM structure can be understood from a simple phenomenological argument – the tunneling and subsequent e-e scattering occur on an extremely fast femtosecond time scale, which leaves very little time for the SPP excitation to take place. Hence, any configuration in which the electron's transport is impeded by elastic collisions, such as reflections, would increase the probability of SPP emission.

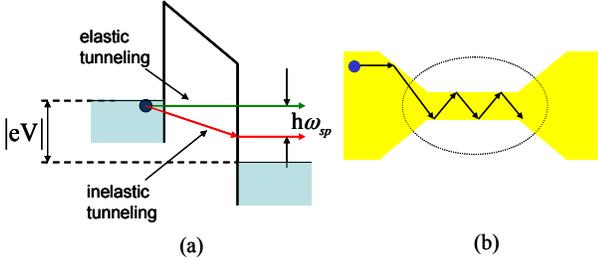

**Figure 1.** (a) Energy diagram of the MIM structure: $|eV|$ is the difference between the Fermi-levels of left and right metal electrodes. Electrons may tunnel between the electrodes elastically (green arrow) or inelastically (red arrows) with emission of an SPP with energy $\hbar\omega_{sp}$. (b) Illustration of multiple collisions of electron with walls of the constriction.

In accordance with this intuitive understanding, in this *paper* we are suggesting to use nanoscale mesoscopic contact, illustrated in Fig 2a as the *active element* of the optical antenna used to couple electrons to the SPP modes of the structures. Similarly to tunneling structures discussed above, the voltage $V$ applied across the constriction falls over the contact of nanoscale length. Electrons passing through the contact quasi-*ballistically* gain the energy $eV$ and can, in principle, transfer it to excited SPP mode, provided that both energy and momentum conservation is maintained. Below we describe mechanisms of plasmonic excitation by electrons in the constriction, and show that with such strategy one can reach QE of plasmonic excitation much higher (by several orders) than with conventional tunneling structures.

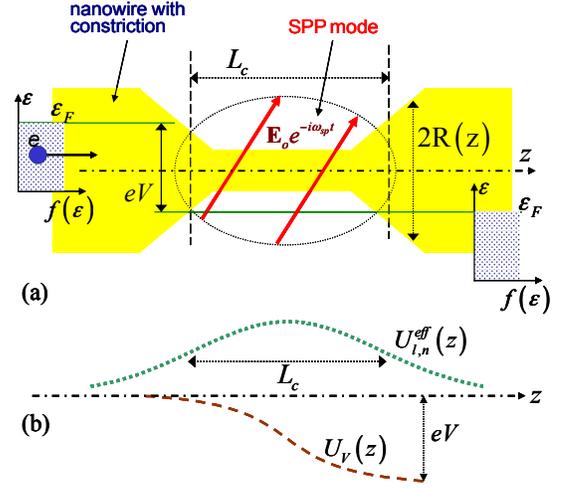

Figure 2. (a): Cylindrical nanowire with constriction; SSP mode (pink) is outlined with dash line, $\mathbf{E}_o \cdot \exp(-i\omega_{sp}t)$ is the field of the mode. (b): The behavior of the potential $U_V(z)$ when the voltage $V$ is applied (dashed), and of the effective barrier along the constriction (dotted).

A proposal for a light source utilizing light emission by electron, passing ballistically through nanoconstriction, has been first made in ref.[20] where the first order-of-magnitude estimates were made using an oversimplified 2D model. In the present paper we develop (and justify in Supporting Information) a rigorous full 3D model for nanowire with nanoscale constriction that allows us to get new results, particularly for transversely polarized waves.

## MODEL

Fig. 2a illustrates a cylindrical nanowire with constriction, characterized by the radius $R(z)$ changing slowly along the axis $z$; the minimal radius in the constriction is $R_{min}$ (at $z=0$), the characteristic length of the constriction is $L_c$. In order to provide a quasi-ballistic transport the characteristic sizes $R_{min}$ and $L_c$ must be less than the electron free path length in the material $l_e$: $R_{min}, L_c \leq l_e$. Note that in plasmonic metals $l_e \sim 50\,\text{nm}$. For other requirements on the sizes, see below. The bias voltage $V$ applied to the structure is dropped along the constriction. This leads to the electron potential energy $U_V(z)$ changing along $z$-axis as illustrated by the dashed line in Fig. 2b, similar to the potential in tunneling structures. The Fermi level to right of the constriction is lower by $|eV|$ compared to the Fermi level of the left (see insets in Fig. 2a, illustrating the distributions $f(\varepsilon)$ of electron over their energy $\varepsilon$). This establishes a current in the constriction[21–23], resembling the electronic transport in a conventional tunneling metal-insulator-metal structure. Let us assume that the metallic structure is engineered to support an SPP mode localized near the constriction as outlined by the dashed line in Fig. 2a. The SPP mode $\mathbf{E}_o \exp(-i\omega_{sp}t)$ is characterized by its resonant frequency $\omega_{sp}$ (in the visible – near IR range for noble metals), its effective length is $L_{mode}^z$ along axis $z$, and the mode volume $V_{mode} \sim L_{mode}^z \cdot S_{mode}$ where $S_{mode}$ is the effective cross-section of the mode. The passage of the electrons through the constriction is assumed quasi-*ballistic*: the electron does not experience any collisions *inside* the constriction, but can collide with the walls. The electron's motion is governed only by the potential walls of the constriction, the applied voltage $V$ and the field of the mode. Note that electronic transport (not optical properties) in *mesoscopic* nanostructures similar to shown in Fig.1 is under studies since 90th of last century – see, for instance, ref.[21-23].

Similarly to MIM structures, an electron passing the constriction can excite the plasmonic mode if $\hbar\omega_{sp} < |eV|$. Below we are estimating the probability to emit the quantum $\hbar\omega_{sp}$ into the mode by a *single* electron passing through the constriction. In the calculations, we are using an following approach based on Einstein relations between spontaneous and stimulated emission rates[24,16]. First, solving Schrödinger equation (see below and Supporting Information) we find the probability $p^{stim}$ of *stimulated* emission into the mode

$$p^{stim} = B \cdot |\mathbf{E}_o|^2 \qquad (1)$$

where $\mathbf{E}_o$ is the amplitude of the field in the mode, and $B$ is the coefficient, proportional to Einstein's coefficient for stimulated emission and containing the square of the matrix element of the interaction Hamiltonian. Then, using the relationship between stimulated and spontaneous emission (an analog of Einstein's relationship), we are finding the probability $p^{spon}$ of *spontaneous* emission into the mode, by substituting the amplitude of the "vacuum" field of one photon $E_{vac}$ into Eq. (1) to obtain

$$p^{spon} = B \cdot |E_{vac}|^2 \qquad (2)$$

and $|E_{vac}|^2$ can be estimated (see Supporting Information S6) as

$$|E_{vac}|^2 \sim \hbar\omega_{sp}/(2\varepsilon_o \bar{\varepsilon} V_{mode}) \qquad (3)$$

where $\varepsilon_o$ is the vacuum permittivity and $\bar{\varepsilon}$ is an average dielectric constant in the mode.

The *time-dependent* single electron Schrödinger equation for wavefunction of an electron describing the stimulated emission is

$$i\hbar\frac{\partial\Psi}{\partial t} = \left[\frac{1}{2m}\left(-i\hbar\frac{\partial}{\partial z} - e\left(A_{\omega_{sp}}e^{-i\omega_{sp}t} + c.c.\right)\right)^2 + U_V(z) - \frac{\hbar^2}{2m}\left(\frac{\partial^2}{\partial x^2} + \frac{\partial^2}{\partial y^2}\right) + U(z,\rho) + e\left(\varphi_{\omega_{sp}}e^{-i\omega_{sp}t} + c.c.\right)\right]\cdot\Psi \qquad (4)$$

where the potential $U(z,\rho)$ describes the electron confinement in the nanowire ($\rho$ is the distance from the axis of the nanowire): $U(z,\rho) = 0$ inside the nanowire [$\rho < R(z)$] and $U(z,\rho) = \infty$ outside the nanowire [$\rho > R(z)$], and $A_{\omega_{sp}}$ and $\varphi_{\omega_{sp}}$ are the amplitude of the vector and scalar potential, representing the correspondingly longitudinal and radial components of the electromagnetic field $\mathbf{E}_o \exp(-i\omega_{sp}t)$ in the plasmonic mode. The vector potential expresses the component of the field along nanowire, and the scalar potential describes the component, normal to the nanowire. Procedure to solve Eq.(4) is given in detail in the Supporting Information.

## RESULTS

In absence of the mode field ($A_o = 0$ and $\varphi_o = 0$), and with an *adiabatic separation of coordinates* applicable when the radius $R(z)$ changes slowly with the coordinate $z$, Eq.(4) has *unperturbed* solutions

$$\Psi_{l,n} = \exp(-itu_o/\hbar) \cdot C_{l,n}[R(z)] \cdot J_l\left(s_{l,n}\frac{\rho}{R(z)}\right) \cdot \frac{e^{il\varphi}}{\sqrt{2\pi}}\phi_{l,n}^{(0)}(z) \qquad (5)$$

describing the *stationary* electron state with the energy $u_{l,n}$; $J_l(s)$ is Bessel function of the order $l$; $C_{l,n}$ is the normalization constant; $s_{l,n}$ is the $n^{th}$ nonzero root of the equation $J_l(s) = 0$; the function $\phi_{l,n}^{(0)}(z)$ satisfies the *one-dimensional* stationary Schrödinger equation

$$-\frac{\hbar^2}{2m}\frac{\partial^2 \phi_{ln}^{(0)}}{\partial z^2} + \left[U_{l,n}^{eff}(z) + U_V(z)\right]\cdot\phi_{ln}^{(0)} = u_{l,n}\cdot\phi_{ln}^{(0)} \qquad (6)$$

where

$$U_{l,n}^{eff}(z) = s_{l,n}^2 \cdot \frac{\hbar^2}{2m \cdot [R(z)]^2} \qquad (7)$$

is the *effective* potential (barrier) due to the transverse confinement of the electron in the nanowire. One can say that the wavefunction (5) with Eqs. (6)-(7) describes electron motion in the channel of the nanowire[19], characterized by the radial number $n$ and the angular number $l$, and correspondingly, $U_{ln}^{eff}(z)$ is the barrier in the channel. Thus, the passage of an electron through the constriction is equivalent to the passage (tunneling) of an electron over (through) the one-dimensional barrier of Eq.(7).

One should note that the above description of the electron motion inside the channel assumes perfectly smooth walls. If the walls are not smooth, i.e. have sub-nanoscale surface roughness, the roughness can lead to scattering of electron between the channels [21-22]. The scattering can be characterized by a scattering rate $R_{scat}$. In Supporting Information S7, we give an estimation of $R_{scat}$ which demonstrates that below estimations of emission probabilities are quite applicable, at least, for not very long nanoconstrictions which we consider in this paper.

It is well known that the emission of photons and SPP's by free electrons requires momentum conservation; hence an electron must collide with a "third body". In *planar* MIM structures, the tunneling barrier is the third body, and the emission can be thought as a "bremsstrahlung" process during electron tunnel-ing. The strong localization of SPP's also can work as "third body" promoting electron-field interaction – see detail description of this in ref. [25].

The metal constriction is exceptionally *anisotropic* electronic structure. The role of the "third" body can be played by either the effective barrier $U_{l,n}^{eff}(z)$ (providing momentum conservation in the longitudinal direction) or by the walls of the constriction itself (conserving the transverse momentum). Therefore the mechanism of SPP emission is expected to exhibit strong dependence on the polarization of the SPP field as shown by red arrows in Fig. 2a. Below we consider separately these two longitudinal and transverse polarizations.

*Longitudinal polarization*

If the field is polarized along the constriction (*parallel* to the z-axis, see Fig.2a), a ballistic electron does not change the channel in which it propagates when emitting the quantum $\hbar\omega_{sp}$ into the mode (see Supporting Information, S3). The probability of stimulated emission into the mode for longitudinal polarisation is written approximately as

$$p_{\parallel}^{stim} = \frac{2e^2}{m\hbar^2\omega_{sp}^4}\sqrt{(u_o - \hbar\omega_{sp})u_o} \cdot |E_o|^2 \qquad (7)$$

In the derivation of formula (7), we assumed that the initial electron energy $u_o$ is larger than the maximum of the effective potential $U_{l_o,n_o}^{eff}(z)$ for the channel $(l_o, n_o)$ in which it propagates, and that the electron is not reflected by the potential. In fact, it means that formula (7) with formulas below for other probabilities below gives the quantum efficiencies of emission into the mode. We assumed also that the mode length $L_{mode}^z$ is large, so that one can consider the longitudinal mode field as independent over the z-axis. This means that the emission into the mode by the electron is exclusively due to collision with the effective one-dimensional barrier $U_{l_o,n_o}^{eff}(z)$ in the channel in which the electron propagates. This process can be described as a "bremsstrahlung" on the effective barrier.

As mention above, the probability of spontaneous emission can be found from Eq.(7) by substituting $|E_o|^2 \rightarrow |E_{vac}|^2$ and using Eq.(3) for $|E_{vac}|^2$:

$$p_{\parallel}^{spon} = \frac{e^2\hbar^2}{\varepsilon_o\bar{\varepsilon}\cdot m \cdot (\hbar\omega_{sp})^3 \cdot V_{mode}}\sqrt{(u_o - \hbar\omega_{sp})u_o} \qquad (8)$$

Assuming realistic values of $u_o \sim \varepsilon_F(\text{Fermi}) \sim 5\text{eV}$, $\hbar\omega_{sp} \sim 1\text{eV}$, $\bar{\varepsilon} \sim 1$ and $V_{mode} = (100\text{nm})^3$, the excitation probability is estimated to be in the range of $p_{\parallel}^{spon} \sim 10^{-5}$. Such a low probability, similar to the probability of emission in planar tunnel structures, is caused by the fact that an electron only experiences a *single* collision with the one-dimensional effective barrier.

*Transverse polarization*

In contrast to a longitudinal polarisation, an electron *does* change the channel as result of emission of the quantum $\hbar\omega_{sp}$ when the field of the mode is polarized *perpendicular* to the z-axis of the constriction (see Fig.2a):

$$(l_o, n_o) \rightarrow (l_1 = l_o \pm 1, n_1) \qquad (9)$$

($n_1$ is integer) – see Supporting Information S5. Obviously, the selection rule (9) is due to the dipole character of interaction of electron in nanowire with the transverse field.

The probability of stimulated emission into the mode for transverse polarisation is written as

$$p_{\perp}^{stim} = \left(d_{l_o n_o}^{l_1 n_1}\right)^2 \left(s_{l_o n_o}^2 - s_{l_1 n_1}^2\right) \cdot \frac{\pi^2 e^2 \cdot \tau_{fligth}^2}{2m\hbar\omega_{sp}} \cdot |E_o|^2 \qquad (10)$$

where $\tau_{fligth} = L_{mode}^z/v_o$ is the electron's time of flight time with the velocity $v_o$ through the nanowire of the length $L_{mode}^z$; $d_{l_o n_o}^{l_1 n_1}$ is the dimensionless dipole moment of transition between the channels $(l_o, n_o)$ and $(l_1, n_1)$, see S5. Note the flight time $\tau_{fligth}$ is, in fact, the interaction time of the electron with the mode field. One should note also that the transition (9) for this field polarisation is *resonant* due to the quantization of electron motion in nanowire in transverse direction so that

$$\hbar\omega_{sp} \approx \left(s_{l_o,n_o}^2 - s_{l_1,n_1}^2\right) \cdot \frac{\hbar^2}{2m\bar{R}^2} \qquad (11)$$

($\bar{R}$ is an average radius of constriction where the electron interacts with the mode field), and this resonant transition is broadened due to finiteness of the interaction time $\tau_{fligth}$ – see S5.

Replacing $|E_o|^2$ by $|E_{vac}|^2 \sim \hbar\omega_{sp}/(2\varepsilon_o V_{mode})$ in Eq.(10), we can estimate the probability of spontaneous emission for the transverse polarisation as

$$p_{\perp}^{spon} = \left(d_{l_o n_o}^{l_1 n_1}\right)^2 \left(s_{l_o n_o}^2 - s_{l_1 n_1}^2\right) \cdot \frac{\pi^2 e^2 \cdot \tau_{fligth}^2}{4m\varepsilon_o V_{mode}} \qquad (12)$$

One should note that since the emission into mode with transverse polarisation involves the transitions (9) between the channels (nanowire levels), the transverse size of constriction must be $\sim 1-5$ nm in order to provide emission in the range $\hbar\omega_{sp} \sim 1\text{eV}$.

This huge enhancement of the probability of emission for a transverse polarisation compared to a longitudinal polarisation arises because of the *multiple* collisions the electron undergoes with the walls of the constriction – see illustration of the collisions in Fig. 1b. One can easily obtain that $p_{\perp}^{spon}$ exceeds $p_{\parallel}^{spon}$ by the factor $\sim(\omega_{sp}\tau_{flight})^2$, where $\omega_{sp}\tau_{flight}$ is approximately the number $N_{col}$ of electron collisions with the walls when the charge flies through the constriction. With above characteristics of emitting electron and nanoconstriction, $N_{col} \sim \omega_{sp}\tau_{flight} \sim 10^2 - 10^3$. Namely this increase of the number of collisions leads to a coupling enhancement between the electron and the transversed-polarized electromagnetic field. In whole, this confirms the importance of electron collisions in the photon emission by charge injection in electron-fed optical antennas. If properly designed, we anticipate the light emitted by a ballistic constriction to surpass common nanoscale sources. Let us consider a current of 10 nA flowing through the constriction, that is approximately $6\times10^{10}$ electron.s$^{-1}$. With an estimated efficiency of $10^{-3}$, the ballistic constriction may release about $6\times10^{7}$ photon.sec$^{-1}$. Assuming an emission wavelength peaking at 800 nm, the optical power radiated by the device is about 15 pW, which is for a constriction with a characteristic size of 1 nm$^2$, leads to a power density of a few kW/cm$^{-2}$.

## CONCLUSIONS

We have shown that electrons passing through a nanoscale metal constriction can very efficiently couple their energy to the electromagnetic field. The coupling mechanism is caused by collisions with the effective barrier or with the walls depending on the polarization of the plasmonic mode. Since the electrons may collide multiple times with the walls, the quantum efficiency can be as high as 10%, a significant increase from the efficiency estimated when the electron experiences a single collision with the effective potential. The strong dependence of SPP emission with the field polarisation stems from the strong anisotropy of the electronic properties of a mesoscopic constriction. In whole, our estimation shows that nanoscale metallic constrictions can be promising alternatives for developing efficient electrically driven nanosource of plasmons.

One should note that although we have focused the discussion on metallic mesoscopic structures, it is obvious that in semiconductor nanowires with ballistic electronic transport one can also realize emission of photons of similar type.

## ASSOCIATED CONTENT

**Supporting Information**

The Supporting Information is available free of charge on the ACS Publications website at DOI:

S1. Schrödinger equation for electron in nanowire with constriction, PDF.

S2. Solution for $\Psi^{(0)}$ (zero order perturbation theory), PDF.

S3. Solution for $\Psi^{(1)}$ (first order perturbation theory), PDF.

S4. Longitudinal polarisation ($E_o^\perp = 0$, $E_o^\parallel = E_o \neq 0$). PDF.

S5. Transverse polarisation ($E_o^\perp = E_o \neq 0$, $E_o^\parallel = 0$), PDF.

S6. The vacuum field $E_{vac}$ in the mode, PDF.

## AUTHOR INFORMATION

**Corresponding Author**

*E-mail: alexusk@lebedev.ru

**Note**

The authors declare no competing financial interests.## ACKNOWLEDGMENT

The work of A. U. was financially supported in part by the Government of the Russian Federation (Grant 074-U01) through ITMO Visiting Professorship program. JK is thankful to US ARO for the support. A. U. is thankful to Russian Science Foundation (Grant 17-19-01532) for support. M. B. and A. B. thank the European Research Council, Grant Agreement No. 306772. The work of A. V, I. S., I. P., M. B. and A. B. is done, in part, in the frames of the Joint Research Project PRS CNRS/RFBR (Grant RFBR-17-58-150007).## REFERENCES

(1) Russel, K. J. Visible light from inelastic tunneling. *Nature Photonics* **2015,** 9, 555.

(2) Cohen, M.; Zalevsky, Z.; Shavit, R. Towards integrated nanoplasmonic logic circuitry. *Nanoscale* **2015,** **5,** 5442.

(3) Lambe, J.; McCarthy, S. L. Light Emission from Inelastic Electron Tunneling. *Phys. Rev. Lett.* 1976, **37**, 923.

(4) Gimzewski, J. K.; Reihl, B.; Coombs, J. H.; Schlittler, R. R. Photon emission with the scanning tunneling microscope. *Z. Phys. B – Condensed Matter* 1988, **72**, 497.

(5) Bharadwaj, P.; Bouhelier, A.; Novotny, L. Electrical Excitation of Surface Plasmons. *Phys. Rev. Lett.* 2011, **106**, 226802.

(6) Wang, T.; Boer-Duchemin, E.; Zhang, Y.; Comtet, G.; Dujardin, G. Excitation of propagating surface plasmons with a scanning tunnelling microscope. *Nanotechnology*, 2011, **22**, 175201.

(7) Buret, M.; Uskov, A. V.; Dellinger, J.; Cazier N.; Mennemanteuil M.-M.; Berthelot, J.; Smetanin, I. V.; Protsenko, I. E.; Colas-des-Francs, G.; Bouhelier, A. Spontaneous Hot-Electron Light Emission from Electron-Fed Optical Antennas. *Nano Lett.* 2015, **15**, 5811.

(8) Kern, J.; Kullock, R.; Prangsma, J.; Emmerling, M.; Kamp, M.; Hecht, B. Electrically driven optical antennas. *Nature Photonics* 2015, **9**, 582.

(9) Huang, K. C. Y.; Seo, M-K.; Sarmiento, T.; Huo, Y.; Harris J. S.; Brongersma, M. L. Electrically driven subwavelength optical nanocircuits. *Nature Photonics*, 2014, **8**, 244.

(11) Parzefall, M.; Bharadwaj, P.; Jain, A.; Taniguchi, T.; Watanabe, K.; Novotny, L. Antenna-coupled photon emission from hexagonal boron nitride tunnel junctions. *Nature Nanotech.*, 2015, **10**, 1058.

(12) Vardi, Y.; Cohen-Hoshen, E.; Shalem, G.; Bar-Joseph, I. Fano Resonance in an Electrically Driven Plasmonic Device. *Nano Lett.*, 2016, **16**, 748.

(13) Cazier, N.; Buret, M.; Uskov, A. V.; Markey, L.; Arocas, J.; Colas Des Francs, G.; Bouhelier, A. Electrical excitation of waveguided surface plasmons by a light-emitting tunneling optical gap antenna. *Optics Express* 2016, **24**, 3873.

(14) Le Moal, E.; Marguet, S.; Canneson, D.; Rogez, B.; Boer-Duchemin, E.; Dujardin, G.; Teperik, T. V.; Marinica, D.-C.; Borisov, A. G. Engineering the emission of light from a scanning tunneling microscope using the plasmonic modes of a nanoparticle. *Phys. Rev. B* 2016, **93**, 035418.

(15) Malinowski, T.; Klein, H.,R.; Iazykov, M.; Dumas, Ph. Infrared light emission from nano hot electron gas created in atomic point contacts *Europhysics Lett.* 2016, **114**, 57002.

(16) Uskov, A. V.; Khurgin, J. B.; Protsenko, I. E.; Smetanin, I. V.; Bouhelier A. Excitation of plasmonic nanoantennas by nonresonant and resonant electron tunneling. *Nanoscale* 2016, **8**, 14573.

(17) Bigourdan, F.; Hugonin, J.-P.; Marquier, F.; Sauvan, Ch.; Greffet, J.-J. Nanoantenna for Electrical Generation of Surface Plasmon Polaritons. *Phys. Rev. Lett.* 2016, **116**, 106803.

(18) Freitag, M; Perebeinos, V; Chen J.; Stein, A; Tsang J. C.; Misewich J. A., Martel R.; Avouris Ph. Hot Carrier Electroluminescence from a Single Carbon Nanotube. *Nano Lett.*, 2004, **4**, 1063.

(19) Adam, E; Aguirre, C. M.; Marty, L.; St-Antoine, B. C.; Meunier, F.; Desjardins, P; D. Ménard D.; Martel R. Electroluminescence from Single-Wall Carbon Nanotube Network Transistors. *Nano Lett.*, 2008, **8**, 2351.

(20) Uskov, A. V.; Khurgin, J. B.; Bouhelier, A.; Buret, M.; Protsenko, I. E.; Smetanin, I. V. "Electrically Driven Optical Antennas enabled by Mesoscopic Contacts", *Proc. SPIE*, 2017, **10102**, 1010204.

(21) Datta, S. Electronic Transport in Mesoscopic Systems, Cambridge University Press (1995).5